\documentclass[aps,prl,twocolumn,superscriptaddress]{revtex4-1}
\usepackage{graphicx}
\begin{document}

%\preprint{}

\title{Quasi-static microdroplet production in a capillary trap}


\author{M. Valet}
\affiliation{Sorbonne Universit\'es, UPMC Univ. Paris 06, UMR 8237, Laboratoire Jean Perrin, F-75005, Paris, France}
\author{L.-L. Pontani}
\affiliation{Sorbonne Universit\'es, UPMC Univ. Paris 06, UMR 8237, Laboratoire Jean Perrin, F-75005, Paris, France}
\affiliation{Sorbonne Universit\'es, UPMC Univ. Paris 06, UMR 7588, Institut des Nanosciences de Paris, F-75005, Paris, France}
\author{A. M. Prevost}
\affiliation{Sorbonne Universit\'es, UPMC Univ. Paris 06, UMR 8237, Laboratoire Jean Perrin, F-75005, Paris, France}
\author{E. Wandersman}
\email[]{elie.wandersman@upmc.fr}
\affiliation{Sorbonne Universit\'es, UPMC Univ. Paris 06, UMR 8237, Laboratoire Jean Perrin, F-75005, Paris, France}

%\thanks{}


\date{\today}

\begin{abstract}
We have developed a method to produce aqueous microdroplets in an oil phase, based on the periodic extraction of a pending droplet across the oil/air interface. This interface forms a capillary trap inside which a droplet can be captured and detached. This process is found to be capillary-based and quasi-static. The droplet size and emission rate are independently governed by the injected volume per cycle and the extraction frequency. We find that the minimum droplet diameter is close to the injection glass capillary diameter and that variations in surface tension moderately perturb the droplet size. A theoretical model based on surface energy minimization in the oil/water/air phases was derived and captures the experimental results. This method enables robust, versatile and tunable production of microdroplets at low production rates.  

\end{abstract}

\pacs{}
%\keywords{}

\maketitle

The development of microfluidics over the last three decades has enabled monodisperse droplet production~\cite{zhu2017passive}, at rapid emission rates ($\sim$ 100 Hz to 10 kHz). In a typical - now popular - flow focusing~\cite{Anna03} or T-junction~\cite{Thorsen01} microfluidic chip, the accessible droplet size covers 1 to a few 100 $\mu$m and is tuned by the channel size and flow rates of the dispersed and continuous phases. These techniques allow fast encapsulation of chemicals and can produce microreactors for the high throughput screening of drugs efficiency~\cite{Miller2012} or directed evolution~\cite{Agresti2010}. They have also been successfully used to produce colloids~\cite{Kong13} and microcapsules~\cite{Shah08}. However, they are not well suited when low production rates are required to allow surfactants with slower dynamics to stabilize the droplets. This is the case, for instance, with protein stabilized emulsions~\cite{Beverung1999}, phospholipids, that are used to create droplet interface bilayers~\cite{Leptihn2013}, or particles, that can lead to micro-structured droplets and colloidosomes~\cite{Kutuzov2007}. For these reasons, various \emph{on demand} microfluidic systems have been developed to produce individual microdroplets with tunable rates. They rely on the introduction of additional external forces (\emph{via} an electric field~\cite{he06}, a mechanical excitation~\cite{bransky09,galas09}, a laser beam\cite{wu12}) that destabilize the oil/water interface and trigger the droplet formation. Such microfluidic systems with tunable production rates make it easy to implement droplet-based 3D printers~\cite{Villar2013} that pave the way to the engineering of artificial tissues~\cite{booth17}.\\
\indent We present in this Letter an additional and simple way to produce aqueous droplets by destabilizing the oil/water interface. This method, surprisingly undocumented, is on demand and consists in pulling out a capillary (filled with the aqueous solution) from the oil/air interface (Fig.~1). We show that the process of droplet formation is quasi-static and yields aqueous droplets  in oil of typical size $d$, the inner capillary diameter [20--700] $\mu$m in the low frequency limit ($<$ 1Hz). Using either a \emph{unique} syringe pump, or just using the hydrostatic pressure, we could easily produce droplets with a polydispersity (standard deviation of the radii distribution over the average radius) of about 1\%, \emph{i.e.} comparable to standard microfluidic techniques. Qualitatively, since the oil(\textit{o})/water(\textit{w}) surface tension is lower than the air(\textit{a})/water one  ($\gamma_{ow} <\gamma_{aw}$), an aqueous droplet can be trapped and detached in oil (Figs.~1 b-d). Quantitatively, these findings are confronted to a model that compares interfacial energies of attached and detached droplets. Compared to the above-mentioned methods~\cite{he06,bransky09,wu12}, this technique relies on a softer forcing of the the oil/water interface, which makes it compatible with the use of any material such as charged components or fragile biological material and amenable for a wide range of applications. \\
\indent The experimental setup (Fig.~1a) consists of a syringe pump (KDS Scientific, Legato 270) that imposes the flow of an aqueous solution at a flow rate $Q$ in a glass capillary (inner diameters d=2$R_0$ ranging from 20 to 700 $\mu$m, details in~\cite{SupMat}). The capillary is mounted on a motorized translation stage (M-ILS250 CCL, Newport) to impose a vertical and cyclic up/down motion across the oil/air interface (Fig.~1a, graphical sketch) with an amplitude $\Delta z$ = 2 mm and a constant velocity $v$, ranging from 100 $\mu$m/s to 5 mm/s. A latency time $T_0= 0.45$ s is required to reverse the direction of the translation stage, during which the capillary stays immobile. The total period of the displacement $T=2(T_0+\Delta z/v)$ is measured and is about 1.7 (\textit{resp.} 40) seconds for $v$= 5 (\textit{resp.} 0.1) mm/s. In the first phase of the experiment, the capillary is immersed in the oil container, and an aqueous droplet grows (Fig.~1b). In the second phase, the capillary moves back up (Fig.~1c) and, as discussed below the droplet may (depending on its size) detach (Fig.~1d) and sediment in the oil phase. Note that to test the versatility of the technique, we also imposed the flow through hydrostatic pressure, which makes the whole setup quite inexpensive (\textit{see}~\cite{SupMat} to build such a setup at low cost).

 \begin{figure}[!t]
\centering
 \includegraphics[width=0.42\textwidth]{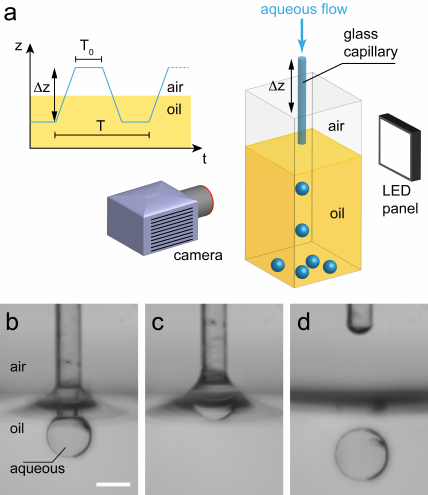}%
 \caption{\label{f1} (a) Sketch of the setup. A syringe pump imposes a flow of water through a glass capillary fixed to a translation stage. This one imposes a displacement of capillary tip $z(t)$ shown in the graphical inset. Imaging is performed with an LED panel and a camera. (b-d) Snapshots obtained with a fast camera. The attached aqueous droplet ([SDS]=8 mM) is immersed in silicon oil at $t=t_0$ (b). The capillary moves up at $t=t_0+123$ ms (c) and the droplet finally detaches at $t=t_0+1.213$ s (d). On (b) the white bar is 400 $\mu$m long.}
 \end{figure}
 
We have characterized the droplet detachment process by optical imaging performed in a transmission geometry, using a LED panel. Two different types of experiments were performed independently. On the one hand, we have used fast imaging with a Photron Fastcam APRX RS camera (1024$\times$1024 pixels$^2$, 8 bits, 3000 fps) to follow the detachment process at short timescales. On the other hand, we have recorded images of hundreds (typically N=100 to 500) of detached droplets for each experimental condition, using a Chameleon3 (Point Grey, 1280$\times$1024 pixels$^2$, 8 bits) camera equipped with a high magnification Navitar objective (maximum spatial resolution of 2.1 $\mu$m/pix) to measure their size distribution. For that purpose, we synchronized the motor displacement and the camera acquisition using Labview (National Instruments). Measures of the droplet radii are obtained by detecting their edges using a custom made MATLAB (MathWorks) routine.\\
To probe the ubiquity of the technique, experiments are performed using both ionic and non-ionic surfactants. As a standard ionic surfactant, we used Sodium Dodecyl Sulfate (SDS) aqueous solutions at various concentrations (from 0.08 mM to the critical micellar concentration~\cite{kanellopoulos} of 8 mM). In this case, the corresponding oil phase is silicon oil (viscosity 5 mPa.s, Sigma Aldrich). As a non-ionic surfactant, we used Span 80 (Sigma Aldrich) dispersed in pure hexadecane at a mass concentration of 2\% (w/w). In that case, the aqueous phase is pure water.

  \begin{figure}[!t]
\centering
 \includegraphics[width=0.45\textwidth]{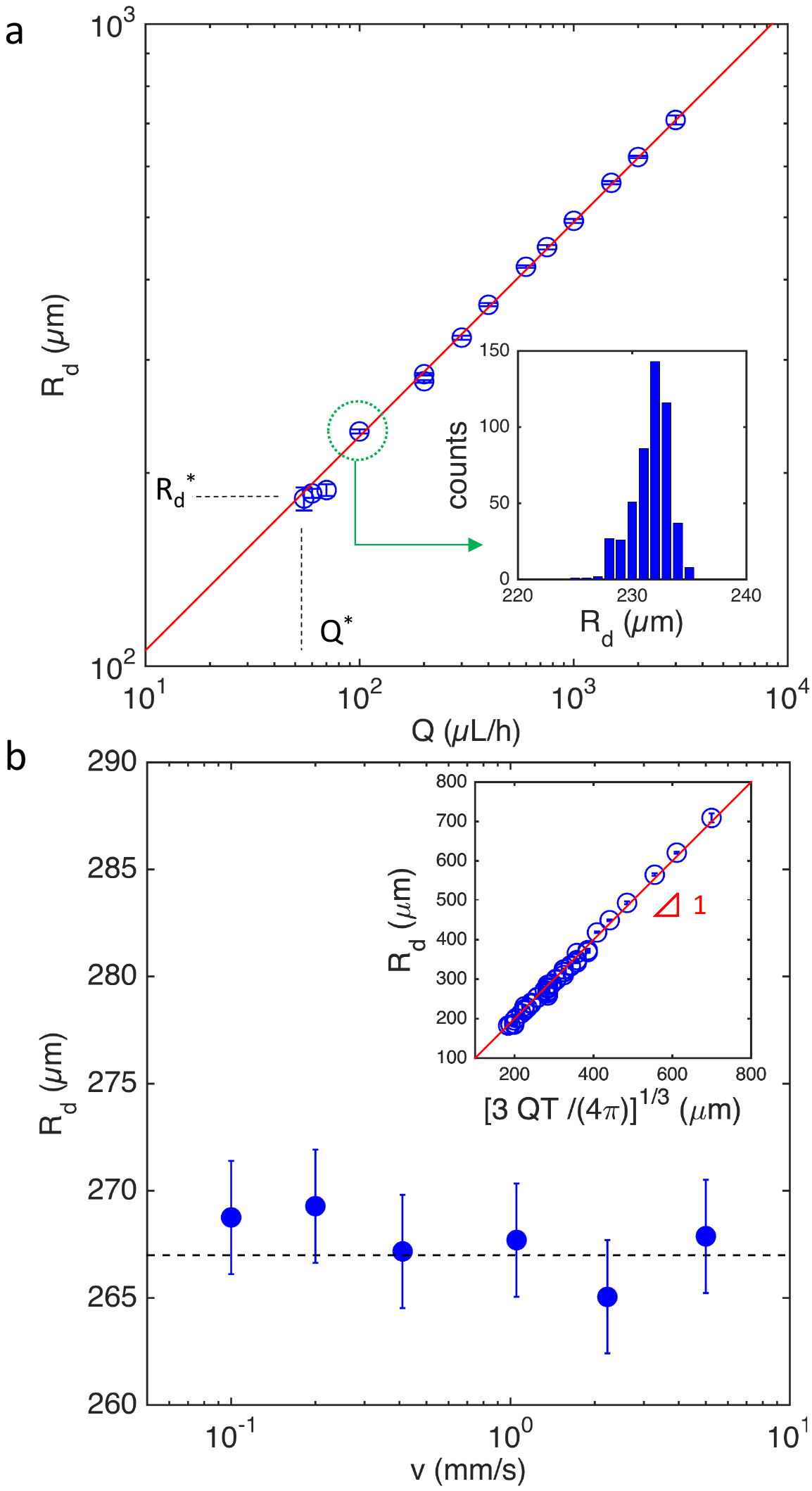}%
 \caption{\label{f2} (a) $R_d$ \textit{versus} $Q$ (d=197 $\mu$m, [SDS] = 8 mM, $v$=5 mm/s). The solid line is a fit $R_d = K Q^{\frac{1}{3}}$, with K = 49.1 $\pm$ 0.3 $\mu$m h$^{1/3}\mu L^{-1/3}$, in good agreement with the expected value K = $(3T/(4\pi))^{1/3}=48.5$. $Q^*$ and the corresponding $R_d^*$ are shown on the graph. Inset: histogram of $R_d$ (N=500 droplets), for $Q=100 \mu$L/h (average radius 232.1 $ \mu$m, standard deviation 1.8 $ \mu$m, yielding a polydispersity of 0.8\%). (b) $R_d$ as a function of $v$, keeping $QT$ constant. From left to right, $Q$ = 8.45, 16.6, 32.6, 75.1, 137.6 and 200 $\mu$L/h, respectively. The dashed line is a guide for the eye. Inset: $R_d$ as function of $[3QT/(4\pi)]^{1/3}$, for all experiments combined, with d=197 $\mu$m (including all $v$ and $[SDS]$ values). The solid line has a slope of 1.}
 \end{figure}
\indent We have first investigated the minimal droplet size one can obtain for a given surface tension. We performed experiments with a capillary ($d$=197 $\mu$m) filled with an SDS solution at 8 mM concentration. We kept the time period of the displacement $T$ constant and decreased gradually the flow rate $Q$. For each $Q$ we measured the average radius (Fig.~2a). For $Q<Q^*$, we do not produce a droplet per cycle. The corresponding minimal droplet radius $R_d^*$ is about 2$R_0$. At a given $Q$, the total volume of injected aqueous phase per cycle is $QT$. One thus trivially expects that $R_d = \left(\frac{3}{4\pi}QT\right)^{1/3}$. The data is well fitted by this equation (Fig.~2a and inset of Fig.~2b). Despite its apparent simplicity, this dependence on $Q$ provides a convenient control parameter to tune the size of the droplet, above $R_d^*$. For droplets larger than $R_d^*$ (by about a factor 1.5), the polydispersity is around 1\% or less (Fig.~2a, inset). We found that approaching the detachment instability limit at $R_d^*$, the polydispersity increases but never exceeds 5\%. Last, note that the droplets produced in the capillary trap have sizes much smaller than what would be obtained by a gravity based destabilization, \emph{i.e.} with an immobile aqueous droplet growing in oil. Using Tate's law~\cite{tate1864}, one expects $R_d^{max} \approx (\frac{3\gamma_{ow}R_0}{2(\rho_w - \rho_o)g})^{1/3} =$ 1.3 mm, with $\rho_{w}$ (\textit{resp.} $\rho_o$) the mass density of water (\textit{resp.} oil). This compares well to our measured value of 1.3 $\pm$ 0.01 mm. The capillary trap method presented in this Letter is therefore efficient to produce small droplets, since $R_d^{max}\approx7 R_d^*$.\\
We have also investigated if the size of the droplet depended on the extraction velocity, by forming droplets at different $v$, in the range [0.1--5] mm/s but keeping $QT$ constant. In practice, we kept $\Delta z$ and $T_0$ constant and adapted the value of $Q$ accordingly. The corresponding radii $R_d$ are plotted as a function of $v$ in Fig.~2b and show that $R_d$ does not depend on $v$. As a consequence, the frequency of the droplet production can be tuned by orders of magnitude, from about 10 mHz to 1 Hz in our case.   
    \begin{figure}[!h]
\centering
 \includegraphics[width=0.5\textwidth]{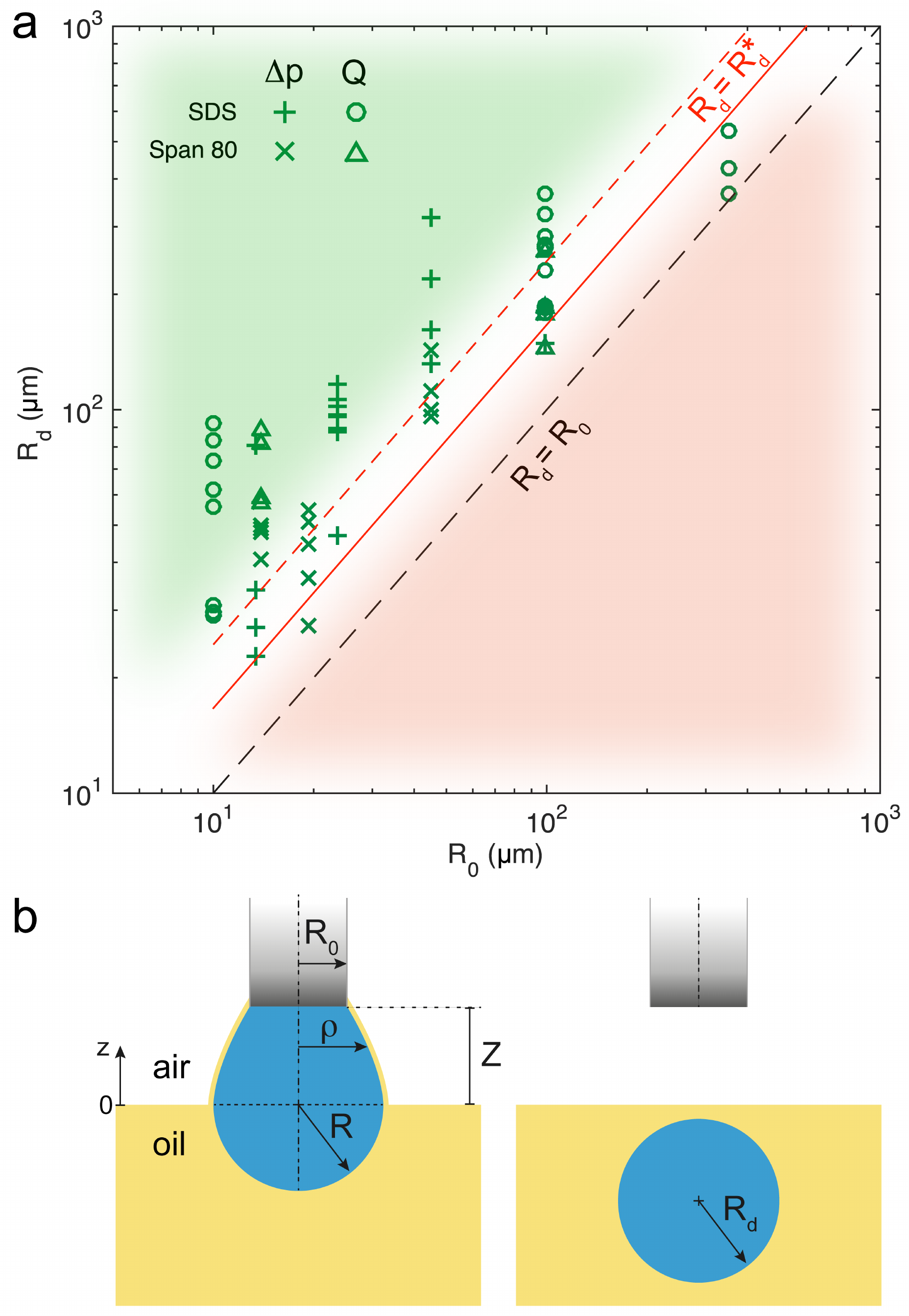}%
 \caption{\label{f3} (a) Phase diagram of the averaged droplet radius $R_d$ as a function of the capillary radius $R_0$. Open (\textit{resp.} cross) symbols corresponds to experiments performed using volume flow rate (resp. hydrostatic pressure) control. The $+$ and $\circ$ (resp. $\times$ and $\triangle$) symbols corresponds to experiments performed at [SDS] = 8~mM concentration (\textit{resp.} Span 80  w/w = 2\%). The solid (\textit{resp.} dashed) red line is the prediction of the model using $\gamma_{ao}$=36 (\textit{resp.} 18)~mN/m. The green (\textit{resp.} red) shaded area corresponds to stable  (\textit{resp.} unaccessible) droplet production. (b) The two states of a droplet in the capillary trap. The pending droplet is composed of a spherical cap of radius $R$ and a paraboloid of revolution, starting from the oil/air free surface at $z=0$ to the capillary tip at $z=Z$. }
 \end{figure}
\indent In a second set of experiments, we have investigated how the size of the droplets depends on $R_0$. On the one hand, we performed these experiments with the  [SDS]=8 mM solution in silicon oil, and on the other hand, with the pure water in Span 80/Hexadecane mixture. Both systems have similar oil/water surface tension ($\gamma_{ow}\approx $ 10 mN/m~\cite{kanellopoulos,Span}). These experiments are performed by either controlling $Q$ or using the hydrostatic pressure to impose the flow. We make sure that the droplet production is stable and only consider experiments for which at least 100 droplets can be produced, with one per cycle. We plot on Fig.~3a the average droplets radii as a function of $R_0$. In this representation, $R_d^*$ is given at a constant $R_0$ by the lowest value in the set of points. Over the whole range of $R_0$'s we obtain $R_d^* \in [R_0-3R_0]$.  For the largest $R_0$=700 $\mu$m, we observe that $R_d^*\approx R_0$. This is likely due to the fact that the size of the droplet approaches $R_d^{max}\sim 1.9$ mm for which gravity effects participate in destabilizing the droplet.\\
\indent We have identified a simple quasi-static mechanism to produce aqueous microdroplets in oil. We have established that its physical  origin is purely capillary, with no $v$ dependent viscous effects. This is expected since the capillary numbers of the problem $\frac{\eta v}{\gamma}$ and $\frac{\eta Q}{\pi R_0^2 \gamma}$ are very small in the explored $v$ and $Q$ ranges. Since the only length scale of the problem is $R_0$, one expects $R_d^*\sim R_0$. To go beyond this scaling argument, we modeled the capillary trap as follows. When fully immersed in oil the pending drop is a spherical cap of radius $R$ attached at the capillary tip. When the capillary tip overpasses the oil/air interface (that defines $z=0$), the aqueous droplet is deformed (Fig.~3b). We postulate that the droplet shape is the union of a spherical cap of radius $R$ and a paraboloid of revolution (radius $\rho(z)\in [R_0-R]$, with $\rho(Z)=R_0$ at the capillary tip). We impose the continuity of the curvature at the cap/paraboloid junction. Considering the smallest unstable droplet, the cap/paraboloid junction has to be located at the sphere equator since it maximizes the droplet surface, which yields $\rho(z)=R+(R_0-R)z^2/Z^2$. For a given droplet of size $R$, volume conservation imposes that $Z/R_0= 10 r^3/(3+4r+8r^2)$ with $r=R/R_0$. Since $r>1$, $Z/R_0\approx 5r/4-5/8$~\cite{Note}. We then deduce the total area of the deformed droplet $A_{tot}$. Close to detachment, fast imaging of the process  (using laser sheet fluorescence imaging~\cite{SupMat}) suggests that a thin film of oil persists and surrounds the droplet. We therefore hypothesize that the oil/air interface follows the paraboloid shape of the aqueous droplet (Fig.~3b, left panel). The model thus neglects the oil meniscus due to the wetting of the glass capillary by the oil phase. Within this geometrical description we can compute the surface free energy in the pending attached state
 \begin{equation}
 F_{a} =  A_{tot} \gamma_{ow} + A_{p} \gamma_{ao} 
 \end{equation}  
 \noindent where $A_{p}$ is the paraboloidal part of the droplet area.
  This energy $F_{att}$ has to be compared to the detached configuration. If the pending droplet is cut at the capillary tip extremity, we produce a detached droplet whose radius $R_d\approx R$~\cite{Note} is simply set by volume conservation. By doing so, we restore an oil/air interface of typical area $\pi R^2 $ and also create an air/water interface at the capillary tip. We therefore write the free energy in the detached state as
   \begin{equation}
 F_{d} =  4\pi R_d^2 \gamma_{ow} + \pi R_0^2 \gamma_{aw} + \pi R^2 \gamma_{ao}
 \end{equation} 
Equating Eqs. (1) and (2), yields
  \begin{equation}
 R_d^* =  \frac{R_0}{2} \left(\frac{5\gamma_{ao}+12 \gamma_{aw}+5\gamma_{ow}}{ 2\gamma_{ao}-\gamma_{ow} }\right)^{1/2}
 \end{equation} 
 
\begin{figure}[!h]
\centering
 \includegraphics[width=0.45\textwidth]{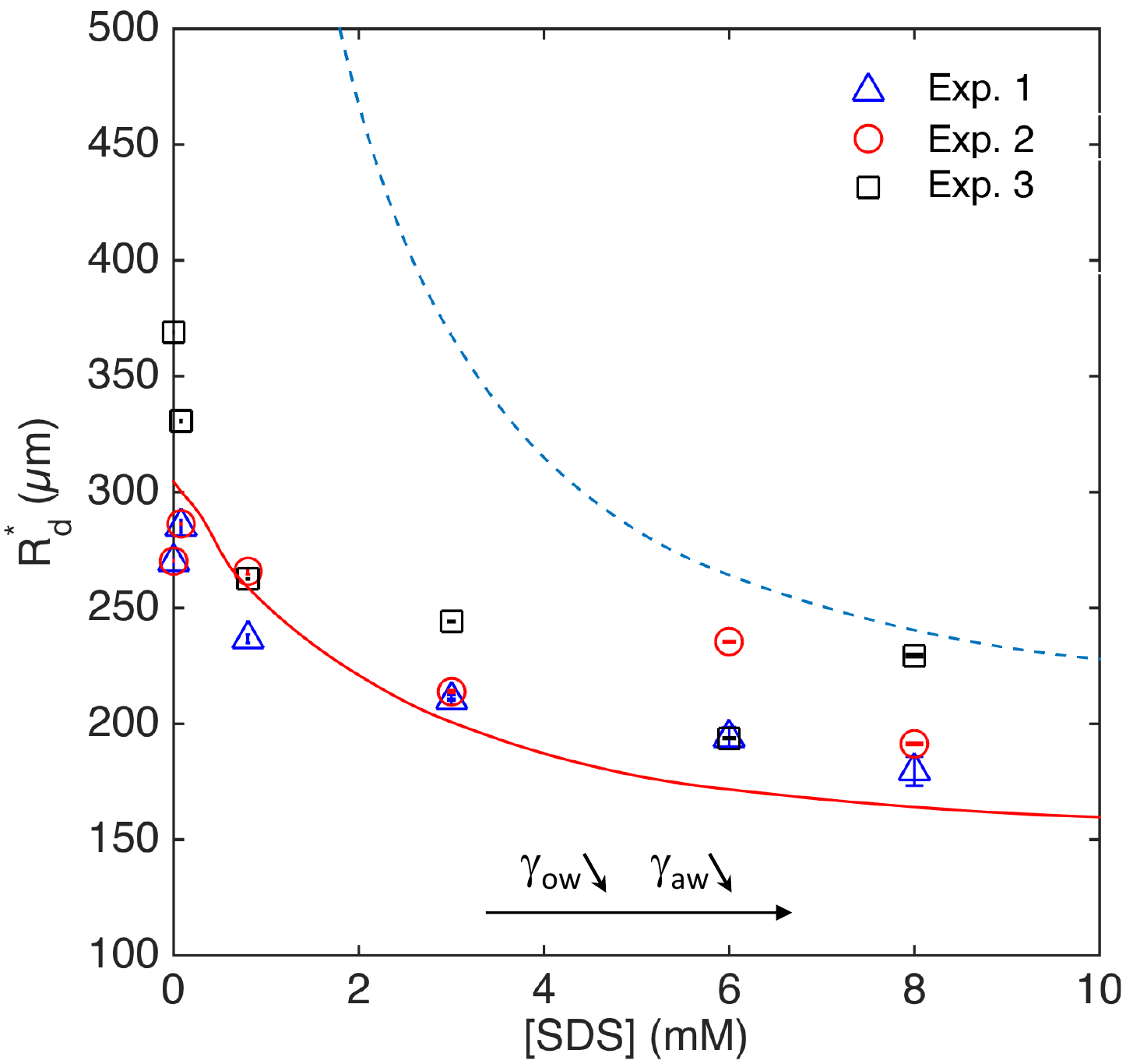}%
 \caption{\label{f3} Minimum radius $R_d^*$ (each point is an average over 200 droplets) as a function of the SDS concentration, for a 197 $\mu$m diameter capillary. For each concentration, the experiment is repeated three times (different symbols) using freshly prepared SDS solutions. The solid (\textit{resp.} dashed) line is the model prediction with $\gamma_{ao}$ = 36 mN/m (\textit{resp.} $\gamma_{ao}$=18 mN/m). }
 \end{figure}
We plot on Fig.~3a the predicted values of $R_d^*$ as a function of $R_0$, using $\gamma_{ow}=11.7$ mN/m~\cite{kanellopoulos} and $\gamma_{aw}=35.9$ mN/m ~\cite{xu17}. We computed $R_d^*$ for $\gamma_{ao}$ in the range [18--36] mN/m, yielding $R_d^*/R_0$ between 2.44 and 1.67, which is in reasonable agreement with the experimental data.\\ 
\indent To further check the validity of our model, we explored the dependence of $R_d^*$ on $\gamma_{ow}$ and $\gamma_{aw}$, by varying the SDS concentration. The results of the experiments ($d$=197 $\mu$m) are presented in Fig.~4, where we plot $R_d^*$ as a function of the SDS concentration. The variation of $R_d^*$ is moderate, increasing of about a factor 1.6 from [SDS]=8 mM to [SDS]=0. As mentioned above, the size measurements close to the instability threshold at $R_d^*$ exhibit stronger fluctuations, which may explain the scattering of the data. The model is derived at each [SDS] concentration using $ \gamma_{ow}$ and $\gamma_{aw} $ deduced from~\cite{kanellopoulos} and~\cite{xu17}. The model captures the slow decrease of $R_d^*$ when the surface tensions $\gamma_{ow}$ and $\gamma_{aw}$ are decreased, as [SDS] increases. At higher $\gamma_{ow}$, Eq.~(3) predicts a divergence of  $R_d^*$ at $\gamma_{ow}=2\gamma_{ao}$. We indeed observe a sharper increase of $R_d^*$ at low [SDS]. The best comparison to our data (Fig.~4) is however obtained with $\gamma_{ao}=36$~mN/m instead of 18~mN/m as measured independently (not shown). Since we neglect in our model the oil meniscus wetting the glass capillary, the oil/air interface area is clearly underestimated.  Taking this point into account should lower the predicted value of $R_d^*$ and may explain this discrepancy with the data. Establishing more precise scaling laws of $R_d^*$ requires additional experimental and theoretical investigations and are beyond the scope of the present Letter.\\
\indent Altogether, our method offers an easy to implement system to produce microdroplets, when low frequencies ($<$1 Hz) are required. It offers an independent control of the size (through the injected volume $QT$) and of the emission frequency. The method is versatile and was found to work with ionic and non ionic surfactants, phospholipids (not shown) and even without any surfactant. We think this method could be particularly interesting in the field of chemical/biological encapsulation, for which precise volume and content control as well as kinetics may be crucial. This method is also very simple to set up and does not require any particular technical facilities~\cite{SupMat}. To increase the emission frequency, parallel droplet production will be implemented. In principle this method should be feasible at smaller scale and could be used to produce droplets size of about 1 $\mu$m.

 \begin{acknowledgments}
We thank Yannick Rondelez, Olivier Dauchot  and Rapha\"{e}l Voituriez for fruitful discussions, Laurence Talini  for surface tension measurements as well as Gaelle el Asmar and Ibra Ndiaye for their help during the experiments.
\end{acknowledgments}


\begin{thebibliography}{22}%
\makeatletter
\providecommand \@ifxundefined [1]{%
 \@ifx{#1\undefined}
}%
\providecommand \@ifnum [1]{%
 \ifnum #1\expandafter \@firstoftwo
 \else \expandafter \@secondoftwo
 \fi
}%
\providecommand \@ifx [1]{%
 \ifx #1\expandafter \@firstoftwo
 \else \expandafter \@secondoftwo
 \fi
}%
\providecommand \natexlab [1]{#1}%
\providecommand \enquote  [1]{``#1''}%
\providecommand \bibnamefont  [1]{#1}%
\providecommand \bibfnamefont [1]{#1}%
\providecommand \citenamefont [1]{#1}%
\providecommand \href@noop [0]{\@secondoftwo}%
\providecommand \href [0]{\begingroup \@sanitize@url \@href}%
\providecommand \@href[1]{\@@startlink{#1}\@@href}%
\providecommand \@@href[1]{\endgroup#1\@@endlink}%
\providecommand \@sanitize@url [0]{\catcode `\\12\catcode `\$12\catcode
  `\&12\catcode `\#12\catcode `\^12\catcode `\_12\catcode `\%12\relax}%
\providecommand \@@startlink[1]{}%
\providecommand \@@endlink[0]{}%
\providecommand \url  [0]{\begingroup\@sanitize@url \@url }%
\providecommand \@url [1]{\endgroup\@href {#1}{\urlprefix }}%
\providecommand \urlprefix  [0]{URL }%
\providecommand \Eprint [0]{\href }%
\providecommand \doibase [0]{http://dx.doi.org/}%
\providecommand \selectlanguage [0]{\@gobble}%
\providecommand \bibinfo  [0]{\@secondoftwo}%
\providecommand \bibfield  [0]{\@secondoftwo}%
\providecommand \translation [1]{[#1]}%
\providecommand \BibitemOpen [0]{}%
\providecommand \bibitemStop [0]{}%
\providecommand \bibitemNoStop [0]{.\EOS\space}%
\providecommand \EOS [0]{\spacefactor3000\relax}%
\providecommand \BibitemShut  [1]{\csname bibitem#1\endcsname}%
\let\auto@bib@innerbib\@empty
%</preamble>
\bibitem [{\citenamefont {Zhu}\ and\ \citenamefont
  {Wang}(2017)}]{zhu2017passive}%
  \BibitemOpen
  \bibfield  {author} {\bibinfo {author} {\bibfnamefont {P.}~\bibnamefont
  {Zhu}}\ and\ \bibinfo {author} {\bibfnamefont {L.}~\bibnamefont {Wang}},\
  }\href@noop {} {\bibfield  {journal} {\bibinfo  {journal} {Lab on a Chip}\
  }\textbf {\bibinfo {volume} {17}},\ \bibinfo {pages} {34} (\bibinfo {year}
  {2017})}\BibitemShut {NoStop}%
\bibitem [{\citenamefont {Anna}\ \emph {et~al.}(2003)\citenamefont {Anna},
  \citenamefont {Bontoux},\ and\ \citenamefont {Stone}}]{Anna03}%
  \BibitemOpen
  \bibfield  {author} {\bibinfo {author} {\bibfnamefont {S.~L.}\ \bibnamefont
  {Anna}}, \bibinfo {author} {\bibfnamefont {N.}~\bibnamefont {Bontoux}}, \
  and\ \bibinfo {author} {\bibfnamefont {H.~A.}\ \bibnamefont {Stone}},\
  }\href@noop {} {\bibfield  {journal} {\bibinfo  {journal} {Applied physics
  letters}\ }\textbf {\bibinfo {volume} {82}},\ \bibinfo {pages} {364}
  (\bibinfo {year} {2003})}\BibitemShut {NoStop}%
\bibitem [{\citenamefont {Thorsen}\ \emph {et~al.}(2001)\citenamefont
  {Thorsen}, \citenamefont {Roberts}, \citenamefont {Arnold},\ and\
  \citenamefont {Quake}}]{Thorsen01}%
  \BibitemOpen
  \bibfield  {author} {\bibinfo {author} {\bibfnamefont {T.}~\bibnamefont
  {Thorsen}}, \bibinfo {author} {\bibfnamefont {R.~W.}\ \bibnamefont
  {Roberts}}, \bibinfo {author} {\bibfnamefont {F.~H.}\ \bibnamefont {Arnold}},
  \ and\ \bibinfo {author} {\bibfnamefont {S.~R.}\ \bibnamefont {Quake}},\
  }\href@noop {} {\bibfield  {journal} {\bibinfo  {journal} {Physical Review
  Letters}\ }\textbf {\bibinfo {volume} {86}},\ \bibinfo {pages} {4163}
  (\bibinfo {year} {2001})}\BibitemShut {NoStop}%
\bibitem [{\citenamefont {Miller}\ \emph {et~al.}(2012)\citenamefont {Miller}
  \emph {et~al.}}]{Miller2012}%
  \BibitemOpen
  \bibfield  {author} {\bibinfo {author} {\bibfnamefont {O.~J.}\ \bibnamefont
  {Miller}} \emph {et~al.},\ }\href {\doibase 10.1073/pnas.1113324109}
  {\bibfield  {journal} {\bibinfo  {journal} {PNAS}\ }\textbf {\bibinfo
  {volume} {109}},\ \bibinfo {pages} {378} (\bibinfo {year}
  {2012})}\BibitemShut {NoStop}%
\bibitem [{\citenamefont {Agresti}\ \emph {et~al.}(2010)\citenamefont {Agresti}
  \emph {et~al.}}]{Agresti2010}%
  \BibitemOpen
  \bibfield  {author} {\bibinfo {author} {\bibfnamefont {J.~J.}\ \bibnamefont
  {Agresti}} \emph {et~al.},\ }\href {\doibase 10.1073/pnas.0910781107}
  {\bibfield  {journal} {\bibinfo  {journal} {PNAS}\ }\textbf {\bibinfo
  {volume} {107}},\ \bibinfo {pages} {4004} (\bibinfo {year}
  {2010})}\BibitemShut {NoStop}%
\bibitem [{\citenamefont {Kong}\ \emph {et~al.}(2013)\citenamefont {Kong} \emph
  {et~al.}}]{Kong13}%
  \BibitemOpen
  \bibfield  {author} {\bibinfo {author} {\bibfnamefont {T.}~\bibnamefont
  {Kong}} \emph {et~al.},\ }\href@noop {} {\bibfield  {journal} {\bibinfo
  {journal} {Soft Matter}\ }\textbf {\bibinfo {volume} {9}},\ \bibinfo {pages}
  {9780} (\bibinfo {year} {2013})}\BibitemShut {NoStop}%
\bibitem [{\citenamefont {Shah}\ \emph {et~al.}(2008)\citenamefont {Shah} \emph
  {et~al.}}]{Shah08}%
  \BibitemOpen
  \bibfield  {author} {\bibinfo {author} {\bibfnamefont {R.~K.}\ \bibnamefont
  {Shah}} \emph {et~al.},\ }\href@noop {} {\bibfield  {journal} {\bibinfo
  {journal} {Soft Matter}\ }\textbf {\bibinfo {volume} {4}},\ \bibinfo {pages}
  {2303} (\bibinfo {year} {2008})}\BibitemShut {NoStop}%
\bibitem [{\citenamefont {Beverung}\ \emph {et~al.}(1999)\citenamefont
  {Beverung}, \citenamefont {Radke},\ and\ \citenamefont
  {Blanch}}]{Beverung1999}%
  \BibitemOpen
  \bibfield  {author} {\bibinfo {author} {\bibfnamefont {C.}~\bibnamefont
  {Beverung}}, \bibinfo {author} {\bibfnamefont {C.}~\bibnamefont {Radke}}, \
  and\ \bibinfo {author} {\bibfnamefont {H.}~\bibnamefont {Blanch}},\ }\href
  {\doibase http://dx.doi.org/10.1016/S0301-4622(99)00082-4} {\bibfield
  {journal} {\bibinfo  {journal} {Biophysical Chemistry}\ }\textbf {\bibinfo
  {volume} {81}},\ \bibinfo {pages} {59 } (\bibinfo {year} {1999})}\BibitemShut
  {NoStop}%
\bibitem [{\citenamefont {Leptihn}\ \emph {et~al.}(2013)\citenamefont {Leptihn}
  \emph {et~al.}}]{Leptihn2013}%
  \BibitemOpen
  \bibfield  {author} {\bibinfo {author} {\bibfnamefont {S.}~\bibnamefont
  {Leptihn}} \emph {et~al.},\ }\href {\doibase 10.1038/nprot.2013.061}
  {\bibfield  {journal} {\bibinfo  {journal} {Nat. Protocols}\ }\textbf
  {\bibinfo {volume} {8}},\ \bibinfo {pages} {1048} (\bibinfo {year}
  {2013})}\BibitemShut {NoStop}%
\bibitem [{\citenamefont {Kutuzov}\ \emph {et~al.}(2007)\citenamefont {Kutuzov}
  \emph {et~al.}}]{Kutuzov2007}%
  \BibitemOpen
  \bibfield  {author} {\bibinfo {author} {\bibfnamefont {S.}~\bibnamefont
  {Kutuzov}} \emph {et~al.},\ }\href {\doibase 10.1039/B710060B} {\bibfield
  {journal} {\bibinfo  {journal} {Phys. Chem. Chem. Phys.}\ }\textbf {\bibinfo
  {volume} {9}},\ \bibinfo {pages} {6351} (\bibinfo {year} {2007})}\BibitemShut
  {NoStop}%
\bibitem [{\citenamefont {He}\ \emph {et~al.}(2006)\citenamefont {He},
  \citenamefont {Kuo},\ and\ \citenamefont {Chiu}}]{he06}%
  \BibitemOpen
  \bibfield  {author} {\bibinfo {author} {\bibfnamefont {M.}~\bibnamefont
  {He}}, \bibinfo {author} {\bibfnamefont {J.~S.}\ \bibnamefont {Kuo}}, \ and\
  \bibinfo {author} {\bibfnamefont {D.~T.}\ \bibnamefont {Chiu}},\ }\href@noop
  {} {\bibfield  {journal} {\bibinfo  {journal} {Langmuir}\ }\textbf {\bibinfo
  {volume} {22}},\ \bibinfo {pages} {6408} (\bibinfo {year}
  {2006})}\BibitemShut {NoStop}%
\bibitem [{\citenamefont {Bransky}\ \emph {et~al.}(2009)\citenamefont
  {Bransky}, \citenamefont {Korin}, \citenamefont {Khoury},\ and\ \citenamefont
  {Levenberg}}]{bransky09}%
  \BibitemOpen
  \bibfield  {author} {\bibinfo {author} {\bibfnamefont {A.}~\bibnamefont
  {Bransky}}, \bibinfo {author} {\bibfnamefont {N.}~\bibnamefont {Korin}},
  \bibinfo {author} {\bibfnamefont {M.}~\bibnamefont {Khoury}}, \ and\ \bibinfo
  {author} {\bibfnamefont {S.}~\bibnamefont {Levenberg}},\ }\href@noop {}
  {\bibfield  {journal} {\bibinfo  {journal} {Lab on a Chip}\ }\textbf
  {\bibinfo {volume} {9}},\ \bibinfo {pages} {516} (\bibinfo {year}
  {2009})}\BibitemShut {NoStop}%
\bibitem [{\citenamefont {Galas}\ \emph {et~al.}(2009)\citenamefont {Galas},
  \citenamefont {Bartolo},\ and\ \citenamefont {Studer}}]{galas09}%
  \BibitemOpen
  \bibfield  {author} {\bibinfo {author} {\bibfnamefont {J.-C.}\ \bibnamefont
  {Galas}}, \bibinfo {author} {\bibfnamefont {D.}~\bibnamefont {Bartolo}}, \
  and\ \bibinfo {author} {\bibfnamefont {V.}~\bibnamefont {Studer}},\
  }\href@noop {} {\bibfield  {journal} {\bibinfo  {journal} {New Journal of
  Physics}\ }\textbf {\bibinfo {volume} {11}},\ \bibinfo {pages} {075027}
  (\bibinfo {year} {2009})}\BibitemShut {NoStop}%
\bibitem [{\citenamefont {Wu}\ \emph {et~al.}(2012)\citenamefont {Wu} \emph
  {et~al.}}]{wu12}%
  \BibitemOpen
  \bibfield  {author} {\bibinfo {author} {\bibfnamefont {T.-H.}\ \bibnamefont
  {Wu}} \emph {et~al.},\ }\href@noop {} {\bibfield  {journal} {\bibinfo
  {journal} {Lab on a Chip}\ }\textbf {\bibinfo {volume} {12}},\ \bibinfo
  {pages} {1378} (\bibinfo {year} {2012})}\BibitemShut {NoStop}%
\bibitem [{\citenamefont {Villar}\ \emph {et~al.}(2013)\citenamefont {Villar},
  \citenamefont {Graham},\ and\ \citenamefont {Bayley}}]{Villar2013}%
  \BibitemOpen
  \bibfield  {author} {\bibinfo {author} {\bibfnamefont {G.}~\bibnamefont
  {Villar}}, \bibinfo {author} {\bibfnamefont {A.~D.}\ \bibnamefont {Graham}},
  \ and\ \bibinfo {author} {\bibfnamefont {H.}~\bibnamefont {Bayley}},\ }\href
  {\doibase 10.1126/science.1229495} {\bibfield  {journal} {\bibinfo  {journal}
  {Science}\ }\textbf {\bibinfo {volume} {340}},\ \bibinfo {pages} {48}
  (\bibinfo {year} {2013})}\BibitemShut {NoStop}%
\bibitem [{\citenamefont {Booth}\ \emph {et~al.}(2017)\citenamefont {Booth},
  \citenamefont {Schild}, \citenamefont {Downs},\ and\ \citenamefont
  {Bayley}}]{booth17}%
  \BibitemOpen
  \bibfield  {author} {\bibinfo {author} {\bibfnamefont {M.}~\bibnamefont
  {Booth}}, \bibinfo {author} {\bibfnamefont {V.~R.}\ \bibnamefont {Schild}},
  \bibinfo {author} {\bibfnamefont {F.}~\bibnamefont {Downs}}, \ and\ \bibinfo
  {author} {\bibfnamefont {H.}~\bibnamefont {Bayley}},\ }\href {\doibase
  10.1039/C7MB00192D} {\bibfield  {journal} {\bibinfo  {journal} {Molecular
  BioSystems}\ ,\ \bibinfo {pages} {published online}} (\bibinfo {year}
  {2017})}\BibitemShut {NoStop}%
\bibitem [{Sup()}]{SupMat}%
  \BibitemOpen
  \href@noop {} {\bibinfo  {journal} {Supplemental Material}\ }\BibitemShut
  {NoStop}%
\bibitem [{\citenamefont {Kanellopoulos}\ and\ \citenamefont
  {Owen}(1971)}]{kanellopoulos}%
  \BibitemOpen
\bibfield  {journal} {  }\bibfield  {author} {\bibinfo {author} {\bibfnamefont
  {A.}~\bibnamefont {Kanellopoulos}}\ and\ \bibinfo {author} {\bibfnamefont
  {M.}~\bibnamefont {Owen}},\ }\href@noop {} {\bibfield  {journal} {\bibinfo
  {journal} {Transactions of the Faraday Society}\ }\textbf {\bibinfo {volume}
  {67}},\ \bibinfo {pages} {3127} (\bibinfo {year} {1971})}\BibitemShut
  {NoStop}%
\bibitem [{\citenamefont {Tate}(1864)}]{tate1864}%
  \BibitemOpen
  \bibfield  {author} {\bibinfo {author} {\bibfnamefont {T.}~\bibnamefont
  {Tate}},\ }\href@noop {} {\bibfield  {journal} {\bibinfo  {journal} {The
  London, Edinburgh, and Dublin Philosophical Magazine and Journal of Science}\
  }\textbf {\bibinfo {volume} {27}},\ \bibinfo {pages} {176} (\bibinfo {year}
  {1864})}\BibitemShut {NoStop}%
\bibitem [{\citenamefont {Cort{\'e}s-Estrada}\ \emph
  {et~al.}(2014)\citenamefont {Cort{\'e}s-Estrada} \emph {et~al.}}]{Span}%
  \BibitemOpen
  \bibfield  {author} {\bibinfo {author} {\bibfnamefont {A.~H.}\ \bibnamefont
  {Cort{\'e}s-Estrada}} \emph {et~al.},\ }\enquote {\bibinfo {title} {Surface
  tension and interfacial tension measurements in water-surfactant-oil systems
  using pendant drop technique},}\ in\ \href@noop {} {\emph {\bibinfo
  {booktitle} {Experimental and Computational Fluid Mechanics}}}\ (\bibinfo
  {publisher} {Springer International Publishing},\ \bibinfo {address} {Cham},\
  \bibinfo {year} {2014})\ pp.\ \bibinfo {pages} {219--226}\BibitemShut
  {NoStop}%
\bibitem [{Not()}]{Note}%
  \BibitemOpen
  \href@noop {} {\bibinfo  {journal} {We checked numerically that this
  approximation has no effect on the model predictions}\ }\BibitemShut
  {NoStop}%
\bibitem [{\citenamefont {Xu}\ \emph {et~al.}(2017)\citenamefont {Xu} \emph
  {et~al.}}]{xu17}%
  \BibitemOpen
\bibfield  {journal} {  }\bibfield  {author} {\bibinfo {author} {\bibfnamefont
  {M.}~\bibnamefont {Xu}} \emph {et~al.},\ }\href@noop {} {\bibfield  {journal}
  {\bibinfo  {journal} {RSC Advances}\ }\textbf {\bibinfo {volume} {7}},\
  \bibinfo {pages} {29742} (\bibinfo {year} {2017})}\BibitemShut {NoStop}%
\end{thebibliography}
\end{document}